\documentstyle[fleqn,12pt]{article}
\def\beq{\begin{equation}}
\def\eeq{\end{equation}}
\def\0{\otimes}
\def\1{\mbox{1\hskip-.25em l}}  
\def\6{\langle}
\def\9{\rangle}
\def\half{\mbox{$1\over2$}}
\def\bk{{\bf k}}
\def\bE{{\bf E}}
\def\bB{{\bf B}}

\def\bep{\mbox{\boldmath $\epsilon$}}
\def\bal{\mbox{\boldmath $\alpha$}}

\def\bb{{\bf b}}
\def\be{{\bf e}}

\def\bq{{\bf q}}
\def\bx{{\bf x}}
\def\by{{\bf y}}
\def\bz{{\bf z}}
\def\hbk{\hat{\bk}}
\def\hbx{\hat{\bx}}
\def\hby{\hat{\by}}
\def\hbz{\hat{\bz}}

\def\bay{\begin{array}}
\def\eay{\end{array}}

\begin{document}
\begin{center}
{\large{\bf Relativistic Doppler Effect in Quantum
Communication}}\\[10mm]
ASHER PERES and DANIEL R. TERNO\\[8mm]
Department of Physics, Technion---Israel Institute of Technology,
32000 Haifa, Israel\end{center}\vskip15mm

\noindent{\bf Abstract. }\ When an electromagnetic signal propagates
in vacuo, a polarization detector cannot be rigorously perpendicular
to the wave vector because of diffraction effects. The vacuum behaves
as a noisy channel, even if the detectors are perfect. The ``noise''
can however be reduced and nearly cancelled by a relative motion of
the observer toward the source. The standard definition of a reduced
density matrix fails for photon polarization, because the transversality
condition behaves like a superselection rule.  We can however define
an effective reduced density matrix which corresponds to a restricted
class of positive operator-valued measures. There are no pure photon
qubits, and no exactly orthogonal qubit states.\vskip15mm

\noindent{\bf1. \ Introduction}

The long range propagation of polarized photons is an essential
tool of quantum crypto\-graphy \cite{gisin}. Usually, optical
fibers are used, and the photons may be absorbed or depolarized
due to imperfections. In some cases, such as communication with
space stations, the photons must propagate in vacuo
\cite{but00}. The beam then has a finite diffraction angle of order
$\lambda/a$, where $a$ is the aperture size, and new deleterious
effects appear. In particular a polarization detector cannot be
rigorously perpendicular to the wave vector, and the transmission is
never faithful, even with perfect detectors. Moreover, the
``vacuum noise'' depends on the relative motion of the observer
with respect to the source.

The relativistic effects reported here are essentially different
from those for massive particles \cite{pst} because massless
particles have only two linearly independent polarization states.
The properties that we discuss are kinematical, not dynamical. At
the statistical level, it is not even necessary to involve quantum
electrodynamics. Most formulas can be derived by elementary
classical methods as shown below. It is only when we need to
consider individual photons, for crypto\-graphic applications,
that quantum theory becomes essential.

This article consists of two parts. First we consider the
propagation of a classical electromagnetic wave. The wave vector
cannot be constant because of diffraction effects.  A polarization
detector cannot unambiguously distinguish orthogonal
polarizations, even if the detector is perfect.  The vacuum
behaves as a noisy channel. We then show that this ``noise'' can be
reduced and nearly cancelled by a relative motion of the observer
toward the source.

In the second part of this paper, we investigate the transmission
of a single photon. The diffraction effects mentioned above lead to
superselection rules which make it impossible to define a reduced
density matrix for polarization. It is still possible to have
``effective'' density matrices; however, the latter depend not only
on the preparation process, but also on the type of detection that is
used by the observer.  \bigskip

\noindent{\bf2. \ Classical electromagnetic signals}

Assume for simplicity that the electromagnetic signal is
monochromatic. In a Fourier decomposition, the Cartesian
components of the wave vector $k_\mu$ (with $\mu=0,1,2,3$) can be
written in term of polar angles:
\beq
k_\mu=(1,\sin\theta\cos\phi,\sin\theta\sin\phi,\cos\theta),
\eeq
where we use units such that $c=1$ and $k_0=1$. Let us choose the
$z$ axis so that a well collimated beam has a large amplitude only
for small $\theta$, and let us rotate the $x$ and $y$ axes so that a
particular
\bk\ in which we are interested has $\phi=0$ (we shall later return to
arbitrary \bk\ with $\phi\neq0$). The Fourier transform of the
electric field is perpendicular to
\beq
\bk=(\sin\theta,0,\cos\theta).
\eeq
If the emitter (``Alice'') selects a linear polarization angle
$a$, the Fourier transform of \bE\ is proportional to
\beq
\bE_\bk=(\cos a\cos\theta,\sin a,-\cos a\sin\theta).
\eeq
The magnetic field Fourier transform is proportional to
$\bB_\bk=\bk\times\bE_\bk$. The Poynting vector {\bf P} is
parallel to \bk\ and gives the energy flux, as usual. We shall
henceforth omit the subscript~\bk.

Suppose that the receiver (``Bob'') has an infinite flat detector
parallel to the $xy$ plane. Then only the component
\beq
P_z\propto E_x^2+E_y^2=\cos^2a\cos^2\theta+\sin^2a
\eeq
is absorbed by the detector. Moreover, if Bob selects a
polarization angle $b$ in the $xy$ plane, the flux detected by him
will be proportional to
\beq
(E_x\cos b+E_y\sin b)^2=(\cos a\cos b\cos\theta+\sin a\sin b)^2.
\label{intensity}
 \eeq
Under ideal conditions ($\theta=0$), the maximal signal behaves as
$\cos^2(a-b)$. For small but finite $\theta$, the fraction of the
signal that is lost, when $a=b$, is $\theta^2\cos^2a$. More
generally it is $\theta^2\cos^2(a-\phi)$, where $\phi$ is the
azimuthal angle of \bk, which was set to zero in the preceding
calculation by a suitable choice of the $x$ axis.

Apart from the above loss in intensity, the angle mismatch may
introduce errors. Consider again  the case $\phi=0$. When Alice
emits a wave that is linearly polarized along the $x$ or $y$ axes,
Bob's intensity pattern varies as $A_x^2 \cos^2 b$ and $A_y^2
\sin^2 b$, respectively. These intensity distributions would be
unambiguously recognized as pertaining to the corresponding linear
polarizations. However, for a general linear polarization state
that Alice may send ($0\neq a\neq\pi/2$) the intensity will be
distributed according to Eq.~(\ref{intensity}). As a result, the
angle $a'$ that Bob will ascribe to it would be related to the
actual polarization direction by $\tan a'=\tan a/\cos\theta$.

In a real experiment, the angles $\theta$ and $\phi$ are
distributed in a continuous way around the $z$ axis (exactly how
depends on the properties of the laser) and one has to take a
suitable average over them. Since the definition of polarization
explicitly depends on the direction of \bk,  taking the average
over many values of \bk\ leads to an impure polarization and
therefore may cause not only an attenuation of the beam, but also
identification errors.

Let us now consider the effect of a motion of Bob relative to
Alice, with a constant velocity ${\bf v}=(0,0,v)$. The Lorentz
transformation of $k_\mu$ in Eq.~(1) yields new components
\beq
k'_0=\gamma(1-v\cos\theta)\qquad\mbox{and}\qquad
  k'_z=\gamma(\cos\theta-v),  \label{gamma}
  \eeq
where $\gamma=(1-v^2)^{-1/2}$. Considering again a single Fourier
component, we have, instead of the unit vector \bk, a new unit
vector
\beq
\bk'=\left(\frac{\sin\theta}{\gamma(1-v\cos\theta)},\;0,\;
  \frac{\cos\theta-v}{1-v\cos\theta}\right).\label{loren1}
  \eeq
In other words, there is a new tilt angle $\theta'$ given by
\beq
\sin\theta'=\sin\theta/\gamma(1-v\cos\theta).
\eeq
For small $\theta$, such that $\theta^2\ll|v|$, we have
\beq
\theta'=\theta \sqrt{{1+v}\over{1-v}}. \label{doppler}
\eeq
The square root is the familiar relativistic Doppler factor. For
large negative $v$, the diffraction angle becomes arbitrarily
small, and sideway losses can be reduced to zero.

It is noteworthy that the same Doppler factor was obtained by Jarett and
Cover \cite{cover} who considered only the relativistic transformations
of bit rate and noise intensity, without any specific physical model.
This remarkable agreement shows that information theory should properly
be considered as a branch of physics.  \bigskip

\noindent{\bf3. \ Quantized electromagnetic signals}

We now turn to the quantum language in order to discuss
applications to secure communication. The ideal scenario is that
Alice sends isolated photons (one particle Fock states). In a more
realistic setup, the transmission is by means of weak coherent
pulses containing on the average less than one photon each.

A basis of the one-photon space is spanned by states of definite
momentum and helicity,
\beq |\bk,\bep_\bk^\pm\9 \equiv |\bk\9\otimes|\bep_\bk^\pm\9,
 \label{basis}\eeq
where helicity states $|\bep_\bk^\pm\9$ are explicitly defined in
Eq.~(\ref{helivectors}) below. The momentum basis is normalized by
\beq
\6\bq|\bk\9=(2\pi)^3(2 k^0)\delta^{(3)}(\bq-\bk).
\eeq

Polarization states that correspond to different momenta belong to
distinct Hilbert spaces and cannot be superposed (an expression
such as $|\bep_\bk^\pm\9+|\bep_\bq^\pm\9$ is meaningless if
$\bk\neq\bq$). The complete basis (\ref{basis}) does not violate this
superselection rule, owing to the othogonality of the momentum
basis. Therefore, a generic one-photon state is given by a wave
packet~\cite{wolf}
\beq
|\Psi\9=\int d\mu(\bk)f(\bk)|\bk,\bal(\bk)\9,\label{photon}
\eeq
where the polarization state $|\bal(\bk)\9$ corresponds to the
3-vector
\beq
\bal(\bk)=\alpha_+(\bk)\bep^+_\bk+\alpha_-(\bk)\bep^-_\bk,
 \label{elliptic}\eeq
$|\alpha_+|^2+|\alpha_-|^2=1$, and the explicit form of
$\bep^\pm_\bk$ is given below. The Lorentz--invariant measure is
\beq
d\mu(k)=\frac{d^3\bk}{(2\pi)^3 2k^0},
\eeq
and normalized states satisfy $\int d\mu(k)|f(\bk)|^2=1$.  Since
diffraction angles and frequency spreads are usually small
\cite{but00, wolf}, $f(\bk)$ significantly differs from zero only
in the vicinity of a certain momentum that we shall denote by
$\bk_A$.

Lorentz transformations of quantum states are most easily computed
by referring to some standard momentum, which for photons is
$p^\nu=(1,0,0,1)$. Accordingly, standard right and left circular
polarization vectors are $\bep^\pm_p=(1,\pm i,0)/\sqrt{2}$. If we
are interested in linear polarization, all we have to do is to use
Eq.~(\ref{elliptic}) with $\alpha_+=(\alpha_-)^*$, so that
$\bal(\bk)$ is real. In general, $\bal(\bk)$ corresponds to
elliptic polarization.

Under a Lorentz transformation $\Lambda$, these states become
$|\bk_\Lambda,\bal(\bk_\Lambda)\9$, where $\bk_\Lambda$ is the
spatial part of a four-vector $k_\Lambda=\Lambda k$, and the new
polarization vector can be obtained by an appropriate rotation
that is described below.  For each $\bk$ a polarization basis is
labeled by the helicity vectors,
\beq
\bep^\pm_\bk=R(\hbk)\bep^\pm_p. \label{helivectors}
\eeq
and the corresponding quantum states are just
$|\bk,\bep^\pm_\bk\9$.
As usual, $\hbk$ denotes the unit 3-vector in the direction of
$\bk$. The matrix that rotates the standard direction $(0,0,1)$ to
$\hbk=(\sin\theta\cos\phi,\sin\theta\sin\phi,\cos\theta)$  is
\beq
R(\hbk)=\left(\bay{ccccc}
\cos\theta\cos\phi & & -\sin\phi  & & \cos\phi\sin\theta \\
\cos\theta\sin\phi & & \cos\phi & & \sin\phi\sin\theta \\
-\sin\theta & & 0 & & \cos\theta
\eay\right),
\eeq
and likewise for $\hbk_\Lambda$.

Under a general Lorentz transformation, be it a rotation or a
boost, helicity is preserved, but the states (and corresponding
geometric vectors)  acquire helicity-dependent phases,
\beq
\alpha_+\bep^+_\bk+\alpha_-\bep^-_\bk\rightarrow \alpha_+e^{i\xi(\Lambda,\hbk)}\bep^+_{\bk_\Lambda}+
\alpha_-e^{-i\xi(\Lambda,\hbk)}\bep^-_{\bk_\Lambda},
\eeq
where the explicit expressions for $\xi(\Lambda,\hbk)$ are given
in
\cite{lpt} The superselection rule that was mentioned above makes it
impossible to define a reduced density matrix in the usual way. We
can however define an ``effective'' reduced density matrix for
polarization, as follows. The labelling of polarization states by
Euclidean vectors $\be_\bk^n$, and the fact that photons are
spin-1 particles, suggest the use of a $3\times 3$ matrix with
entries labelled $x$, $y$ and $z$. Classically, they correspond to
different directions of the electric field. For example, when
$\bk=k_A\hat{\bf z}$, only $\rho_{xx}$, $\rho_{xy}$, $\rho_{yy}$
are non-zero.

For a generic photon state $|\Psi\9$, let us try to construct a
reduced density matrix $\rho_{xx}$ that gives the expectation
value of an operator representing the polarization in the $x$
direction, irrespective of the particle's momentum. To have a
momentum-independent polarization is to tacitly admit longitudinal
photons. Therefore, in terms of real transversal photons of a
momentum $\bk$, this problem suggests to find the direction that
is perpendicular to $\bk$ and closest to an arbitrary unit vector
$\hbx$. That is, we are looking for a unit Euclidean complex
vector $\be_x(\bk)$ such that $\bk\cdot\be_x(\bk)=0$ and
$\hbx\cdot\be_x(\bk)=\max$.

Momentum-independent polarization states thus consist of physical
(transversal) and unphysical (longitudinal) parts with a polarization
vector $\bep^\ell=\hbk$. For example, a
polarization state along the $x$-axis is
\beq
|\hbx\9=x_+(\bk)|\bep^+_\bk\9+x_-(\bk)|\bep^-_\bk\9+
x_\ell(\bk)|\bep^\ell_\bk\9,\label{decomp}
\eeq
where $x_\pm(\bk)=\bep^\pm_\bk\cdot\hbx$, and $x_\ell(\bk)=
\hbx\cdot\hbk=\sin\theta\cos\phi$. It follows that
$|x_+|^2+|x_-|^2+|x_\ell|^2=1$, and
\beq
\be_x(\bk)=\frac{x_+(\bk)\bep^+_\bk+x_-(\bk)\bep^-_\bk}
{\sqrt{x_+^2+x_-^2}} \label{physdir}.
\eeq
Note that $\6\hbx|\hby\9=\hbx\cdot\hby=0$, whence
\beq |\hbx\9\6\hbx|+|\hby\9\6\hby|+|\hbz\9\6\hbz|=\1\label{xyz}. \eeq

To the direction $\hbx$ corresponds a projection operator
\beq
P_{xx}=|\hbx\9\6\hbx|\otimes \1_p=|\hbx\9\6\hbx|\otimes \int
d\mu(\bk)|\bk\9\6\bk|,
\eeq
where $\1_p$ is the unit operator in momentum space. The action of
$P_{xx}$ on $|\Psi\9$ follows from Eq.~(\ref{decomp}) and
$\6\bep^\pm_\bk|\bep^\ell_\bk\9=0$. Only the transversal
part of $|\hbx\9$ appears in the expectation value:
\beq
\6\Psi|P_{xx}|\Psi\9=\int d\mu(\bk)|f(\bk)|^2|x_
 +(\bk)\alpha_+^*(\bk)+x_-(\bk)\alpha_-^*(\bk)|^2.
\eeq
Define the transversal part of $|\hbx\9$:
\beq
|\bb_x(\bk)\9\equiv
 (|\bep^+_\bk\9\6\bep^+_\bk|+|\bep^-_\bk\9\6\bep^-_\bk|)|\hbx\9=
 x_+(\bk)|\bep^+_\bk\9+x_-(\bk)|\bep^-_\bk\9.
\label{vector}
\eeq
Likewise define  $|\bb_y(\bk)\9$ and
$|\bb_z(\bk)\9$. These three state vectors are neither of
unit length nor mutually orthogonal. For $\bk=
(\sin\theta\cos\phi,\sin\theta\sin\phi,\cos\theta)$ we have
\begin{eqnarray}
|\bb_x(\bk)\9 & = & [(\cos\theta\cos\phi+i\sin\phi)|\bep^+_\bk\9+
 (\cos\theta\cos\phi-i\sin\phi)|\bep^-_\bk\9]/\sqrt{2},\\
 {} & = & c(\theta,\phi)|\bk,\be_x(\bk)\9,
\end{eqnarray}
where $\be_x(\bk)$ is given by Eq.~(\ref{physdir}), and
$c(\theta,\phi)=\sqrt{x_+^2+x_-^2}$.

Finally, a POVM element $E_{xx}$ which is the physical part of
$P_{xx}$, namely is equivalent to $P_{xx}$ for physical states
(without longitudinal photons) is
\beq
E_{xx}=\int d\mu(\bk)|\bk,\bb_x(\bk)\9\6\bk,\bb_x(\bk)|,
\eeq
and likewise for other directions. The operators $E_{xx}$, $E_{yy}$
and $E_{zz}$ indeed form a POVM in the space of physical states,
owing to Eq.~(\ref{xyz}). It then follows from Eq.~(\ref{vector})
and similar definitions for other directions that, for any $\bk$,

\beq
|\bb_x(\bk)\9\6\bb_x(\bk)|+|\bb_y(\bk)\9\6\bb_y(\bk)|+
  |\bb_z(\bk)\9\6\bb_z(\bk)|={\1}_{\perp\bk},
\eeq
where ${\1}_{\perp\bk}$ is the identity operator in the subspace
of polarizations orthogonal to $\bk$.

To complete the construction of the density matrix, we introduce
additional directions. Following a standard practice of state
reconstruction \cite{nc97}, we consider $P_{x+z,x+z}$,
$P_{x-iz,x-iz}$ and similar combinations. For example,
\beq
P_{x+z,x+z}=\half(|\hbx\9+|\hbz\9)(\6\hbx|+\6\hbz|)\otimes\1_p
\eeq
The diagonal elements of the new polarization density matrix are
defined as
\beq
\rho_{mm}=\6 \Psi|E_{mm}|\Psi\9,\qquad m=x,y,z,
\eeq
and the off-diagonal elements are recovered by combinations such as
\beq
\rho_{xz}=\6\Psi(|\hbx\9\6\hbz|\otimes\1_p)|\Psi\9=
\6\Psi|E_{x+z,x+z}+E_{x-iz,x-iz}-E_{xx}-E_{zz}|\Psi\9,
\eeq
where we denote $|\hbx\9\6\hbz|\otimes\1_p$ as $P_{xz}$, and by
$E_{xz}$ its ``physical" part. We then get a simple expression for
the reduced density matrix corresponding to the polarization state
$|\bal(\bk)\9$:
\beq
\rho_{mn}=\6\Psi|E_{mn}|\Psi\9=
\int d\mu(\bk)|f(\bk)|^2\6\bal(\bk)|\bb_m(\bk)\9\6\bb_n(\bk)|\bal(\bk)\9
, \quad m,n,=x,y,z. \label{reduced}
\eeq

Our basis states $|\bk,\bep_\bk\9$ are direct products of momentum
and polarization. Owing to the transversality requirement
$\bep(\bk)\cdot\bk=0$, they remain direct products under Lorentz
transformations. All the other states have their polarization and
momentum degrees of freedom entangled. As a result, if one is restricted
to polarization measurements as described by the above POVM, there
do not exist two orthogonal polarization states. In general, any
measurement procedure with finite momentum sensitivity will lead to
the errors in identification. This can be seen as follows.

Let two states $|\Phi\9$ and $|\Psi\9$  be of the form in
Eq.~(\ref{photon}). Their reduced polarization density matrices,
$\rho_\Phi$ and $\rho_\Psi$, respectively, are calculated using
Eq.~(\ref{reduced}). Since the states are entangled, the von
Neumann entropies of the reduced density matrices, $S=-{\rm
tr}(\rho\ln\rho)$, are positive \cite{qt}.  Therefore, both
matrices are at least of rank two. Since the overall dimension
is~3, it follows that ${\rm tr}(\rho_\Phi\rho_\Psi)>0$ and these
states are not perfectly distinguishable. An immediate  corollary
is that photon  polarization states {\em cannot be cloned
perfectly}. This is because no-cloning theorem, in its various
versions \cite{noclo}, forbids an exact copying of unknown
non-orthogonal states.

To quantify the distinguishability of a pair of quantum states,
we shall use the simplest criterion, namely the probability of error
$P_E$, defined as follows: an observer receives a single copy of one of
the two known states and performs any operation permitted by quantum
theory, in order to decide which state was supplied. The probability
of a wrong answer for an optimal measurement is \cite{fu:99}
\beq
P_E(\rho_1,\rho_2)=\half+\mbox{$1\over4$}\,{\rm tr}|\rho_1-\rho_2|,
\label{pe} \eeq
where, for any operator $O$, the operator $|O|$ is defined as
$\sqrt{(O^\dag O)}$. As shown below, the distinguishability of
polarization density matrices depends on the observers' motion.
First we present some general considerations and then illustrate
them with a simple example.

Let us take the $z$-axis to coincide with the average direction of
propagation so that the mean photon momentum is $k_A\hbz$.
Typically, the spread in momentum is small, but not necessarily
equal in all directions. Usually the intensity profile of laser
beams has cylindrical symmetry, and we may assume that
$\Delta_x\sim\Delta_y\sim\Delta_r$ where the index $r$ means
radial. We may also assume that $\Delta_r\gg\Delta_z$, since the
momentum spread along the average propagation direction does not
necessarily increase the mixness  of the reduced density matrix,
as we shell see below. We then have
\beq
f(\bk)\propto f_1[(k_z-k_A)/\Delta_z]\,f_2(k_r/\Delta_r).
\eeq
 We  approximate
\beq
\theta\approx\tan\theta\equiv k_r/k_z\approx k_r/k_A.
\label{theta}
\eeq
In  pictorial language, polarization planes for different momenta
are tilted by angles up to $\sim\Delta_r/k_A$, so that we
expect an error probability of the order $\Delta_r^2/k_A^2$. In
the density matrix $\rho_{mn}$ all the elements of the form
$\rho_{mz}$ should vanish when $\Delta_r\to0$. Moreover, if
$\Delta_z\to0$, the non-vanishing $xy$ block goes to the usual
(monochromatic) polarization density matrix,
\beq
\rho_{{\rm pure}}=\left(\bay{ccc}
|\alpha|^2 & \beta & 0\\
\beta^* & 1-|\alpha|^2 &0\\
0 & 0 &0
\eay\right).
\eeq

As an example we consider two states which, if the momentum spread
could be ignored, would be $|k_A\hat{\bf z},\bep^\pm_{k_A\hat{\bf
z}}\9$. To simplify the calculations we assume a Gaussian distribution:
\beq
f(\bk)=Ne^{-(k_z-k_A)^2/2\Delta_z^2}e^{-k_r^2/2\Delta_r^2},
\eeq
where $N$ is a normalization factor and $\Delta_z\ll\Delta_r$.
Moreover, we take the polarization components to be
$\bep^\pm_\bk\equiv R(\hbk)\bep^\pm_p$. That means we have to
analyze  the states
\beq
|\Psi_\pm\9=\int d\mu(\bk) f(\bk)|\bep^\pm_\bk,\bk\9\label{st},
\eeq
where $f(\bk)$ is given above.

We expand $R(\hbk)$ up to second order in $\theta$. Reduced
density matrices are calculated by techniques similar to those for
massive particles \cite{pst}, using rotational symmetry around the
$z$-axis and normalization requirements. The leading order in
$\Omega\equiv\Delta_r/k_A$ gives
\beq
\rho_+=\half(1-\half\Omega^2)\left(\bay{ccc}
1 & -i & 0\\
i & 1 &0\\
0 & 0 &0
\eay\right)+\half\Omega^2\left(\bay{ccc}
0 & 0 & 0\\
0 & 0 & 0\\
0 & 0 & 1
\eay\right),
\eeq
and $\rho_-=\rho_+^*$. At the same level of precision,
\beq
P_E(\rho_+,\rho_-)=\Delta_r^2/4k_A^2.
\eeq
It is interesting to note that an optimal strategy for
distinguishing between these two states is a polarization
measurement in the $xy$-plane. Then the effective $2\times 2$
density matrices are perfectly distinguishable, but there is a
probability $\Omega^2/2$ that no photon be detected at all. The
above result was valid due to the special form of the states
that we chose. Potential errors in the upper $2\times 2$ blocks
were averaged out in the integration over~$\phi$.

We now calculate Bob's reduced density matrix. We again assume that
Bob moves along the $z$-axis with a velocity $v$. Recall that reduced
density matrices  have no transformation law \cite{pst}, (only the
{\it complete\/} density matrix has one) except in the limiting case of
sharp momenta.  The only way to calculate Bob's reduced density matrix
is to transform the complete state, and only then take a partial trace.
According to Eqs.~(\ref{reduced}) and (\ref{st}), reduced density
matrices in both frames are given by the expression
\beq
(\rho_\pm)_{mn}=\int d\mu(\bk)|f(\bk)|^2\6
R(\hbk)\bep_p^\pm|\bb_m(\bk)\9\6\bb_n(\bk)|R(\hbk)\bep_p^\pm\9.
\eeq
Note that pure boosts preserve the orientation of the coordinate
axes in 3-space, and therefore do not affect the indices of
$\rho_{mn}$. The measure $\mu(\bk)$ is Lorentz-invariant and
$\bep_p^\pm$ are constant by definition. Since $f$ is a scalar
function, it transforms as $f'(\bk)=f(\bk_{\Lambda^{-1}})$, where
primes indicate Bob's frame, as in
Eqs.~(\ref{gamma}--\ref{doppler}).  This is the only frame
dependent expression in (\theequation), since the phase factors
$e^{\pm i\xi}$ cancel out for the helicity eigenstates. Therefore,
there are two equivalent methods to calculate Bob's polarization
density matrix. One is to change the argument of $f(\bk)$ to
$\bk_{\Lambda^{-1}}$, and another is to change the argument of the
rotation matrix $R(\hbk)$ to $\hbk_\Lambda$. Using the second
option, we obtain
\beq
(\rho'_\pm)_{mn}=\int d\mu(\bk)|f(\bk)|^2\6
R(\hbk_\Lambda)\bep_p^\pm|\bb_m(\bk)\9
\6\bb_n(\bk)|R(\hbk_\Lambda)\bep_p^\pm\9.
\eeq

A boost along the $z$-axis preserves $k_r$ and $\phi$. On the
other hand, from Eq.~(\ref{loren1}) it follows that
\beq
k'_z \approx k_A\sqrt{\frac{1-v}{1+v}}.
\eeq
Thus, at leading order in $\theta$ we have
 \beq
 \theta'\approx \sqrt{\frac{1+v}{1-v}} k_r/k_A
 \approx \sqrt{\frac{1+v}{1-v}} \theta,
 \eeq
which is substituted into $R(\bk_\Lambda)$. Since everything else in the
integral remains the same, the effect of relative motion is given
by a substitution
\beq
\Omega\to\sqrt{\frac{1+v}{1-v}}\Omega.
\eeq
It follows that
\beq
P'_E=\frac{1+v}{1-v}P_E, \label{dope}
\eeq
which may be either larger or smaller than $P_E$. As expected, we
obtain for one-photon states the same Doppler effect as in the
preceding classical calculations.

Although reduced polarization density matrices have no general
transformation rule, the above results, as well as the analysis of
massive particles \cite{pst}, show that such rules can be derived
for particular classes of experimental procedures. We can then ask
how these effective transformation rules $\rho'=T[\rho]$ fit into
the framework of general state transformations. A general state
transformation $T$ is usually required to be {\it completely
positive\/} (CP), namely \cite{dav:b,krau:b},
\beq
T[\rho]=\sum_i M_i\rho M_i^\dag,
\eeq
where the $M_i$ are bounded operators. It can be proved that
distinguishability, as expressed by natural measures like $P_E$, cannot
be improved by any CP transformation \cite{fu:99}. It is also known
that the CP requirement may fail if there is a prior entanglement
of $\rho$ with another system \cite{pe:94}.  Since from \cite{pst}
and Eq.~(\ref{dope}) it follows that in our case distinguishability
{\it can\/} be improved, we conclude that these transformations are
not completely positive. The reason is that the Lorentz transformation
acts not only on the ``interesting'' polarization variables, but also
on the ``hidden'' momentum variables that we elected to ignore and to
trace out.

This technicality has one important consequence. In quantum
information theory quantum channels are described as completely
positive maps \cite{holevo,amosov,king} that act on qubit states. Qubits
themselves are realized as particles' discrete degrees of freedom.
If relativistic motion is important, then not only does the vacuum
behave as a noisy quantum channel, but the very representation of a
channel by a CP map fails.

\bigskip \noindent{\bf Acknowledgments}

Work by AP was supported by the Gerard Swope Fund and the Fund for
Encouragement of Research. DRT was supported by a grant from the
Technion Graduate School.

\clearpage

\begin{thebibliography}{99}

\bibitem{gisin}{\sc Gisin, N., Ribordy, G., Tittel, W.,} {\rm and}
{\sc Zbinden, H.}, 2002 {\it Rev. Mod. Phys.\/} {\bf74}, 145.

\bibitem{but00}{\sc Buttler, W. T., Hughes, R.J., Lamoreaux, S. K.,
Morgan, G. L., Nordholt, J. E.,} {\rm and} {\sc Peterson, C. G.}, 2000,
{\it Phys. Rev. Lett.\/} {\bf84}, 5652.

\bibitem{pst}{\sc Peres, A., Scudo, P. F.,} {\rm and} {\sc Terno, D.
R.}, 2002, {\it Phys. Rev. Lett.\/} {\bf88}, 230402.

\bibitem{cover}{\sc Jarett, K.} {\rm and} {\sc Cover, T.}, 1981, {\it
IEEE Trans. Info. Theory\/}, {\bf IT-27}, 152.

\bibitem{wolf}{\sc Mandel, L.} {\rm and} {\sc Wolf, E.}, 1995, {\it
Optical Coherence and Quantum Optics\/}
(Cambridge: Cambridge University Press).

\bibitem{lpt} {\sc Lindner, N. H., Peres, A.} {\rm and} {Terno, D.
R.\/},2003, e-print hep-th/0304017.


\bibitem{nc97}{\sc Chuang, I. L.} {\rm and} {\sc Nielsen, M. A.,} 1997,
{\it J. Modern Optics\/} {\bf44}, 2455.

\bibitem{qt}{\sc Peres, A.}, 1995, {\it Quantum Theory: Concepts and
Methods\/} (Dordrecht: Kluwer).

\bibitem{noclo}{\sc Peres, A.}, 2002, e-print quant-ph/0205076.

\bibitem{fu:99}{\sc Fuchs, C. A.} {\rm and} {\sc van de Graaf, J.},
1999, {\it IEEE Trans. Info. Theory\/}, {\bf IT-45}, 1216.

\bibitem{dav:b}{\sc Davies, E. B.}, 1976, {\it Quantum Dynamics of Open
Systems\/} (New York: Academic Press), ch. 8.

\bibitem{krau:b}{\sc Kraus, K.}, 1983, {\it States, Effects and
Operations\/} (Berlin: Springer).

\bibitem{pe:94}{\sc Pechukas, P.}, 1994, {\it Phys. Rev. Lett.\/}
{\bf73}, 1060.

\bibitem{holevo}{\sc Holevo, A. S.}, 1999, {\it Russ. Math. Surveys\/}
{\bf53}, 1295.

\bibitem{amosov}{\sc Amosov, G. G., Holevo, A. S.}, {\rm and} {\sc
Werner, R. F.}, 2000, {\it Probl. Info. Transmission\/} {\bf36}, 305.

\bibitem{king}{\sc King, C.} {\rm and} {\sc Ruskai, M. B.}, 2001, {\it
IEEE Trans. Info. Theory\/}, {\bf IT-47}, 192.

\end{thebibliography}
\end{document}